# MAC: A NOVEL SYSTEMATICALLY MULTILEVEL CACHE REPLACEMENT POLICY FOR PCM MEMORY

Shenchen Ruan, Haixia Wang and Dongsheng Wang

Tsinghua National Laboratory for Information Science and Technology, Tsinghua University, Beijing, China
rsc13@mails.tsinghua.edu.cn, {hx-wang, wds}@tsinghua.edu.cn

## ABSTRACT

*The rapid development of multi-core system and increase of data-intensive application in recent years call for larger main memory. Traditional DRAM memory can increase its capacity by reducing the feature size of storage cell. Now further scaling of DRAM faces great challenge, and the frequent refresh operations of DRAM can bring a lot of energy consumption. As an emerging technology, Phase Change Memory (PCM) is promising to be used as main memory. It draws wide attention due to the advantages of low power consumption, high density and nonvolatility, while it incurs finite endurance and relatively long write latency. To handle the problem of write, optimizing the cache replacement policy to protect dirty cache block is an efficient way. In this paper, we construct a systematically multilevel structure, and based on it propose a novel cache replacement policy called MAC. MAC can effectively reduce write traffic to PCM memory with low hardware overhead. We conduct simulation experiments on GEM5 to evaluate the performances of MAC and other related works. The results show that MAC performs best in reducing the amount of writes (averagely 25.12%) without increasing the program execution time.*

## 1. INTRODUCTION

Traditional computer storage system generally adopts multi-layer structure and uses DRAM as main memory. The development of multi-core system and sharp increase of data-intensive application put forward higher demand to storage system. As for main memory, a normal way to enlarge the memory capacity is integrating more DRAM cells by reducing the feature size of a single cell. Scaling DRAM becomes extremely hard when the size has reached 20 nm [1, 2]. In response to the challenge, searching for novel storage materials turns to be a hotspot in recent years.

PCM (phase change memory) is an emerging storage technology which is likely to take this responsibility. As shown in Table 1 [15], PCM can provide up to 4x more density than DRAM and the feature of nonvolatility eliminates the power consumption of refresh operations. PCM can only tolerate $10^8$~$10^9$ writes. Main memory in actual system will face frequent writes. If we directly use PCM as main memory without protective measures, it can be disabled in hours. In addition, write to PCM costs much more power and latency than read [3, 4, 5]. There are many technologies proposed focus on the problem of write. Among them, optimizing the cache replacement policy of last level cache (LLC) is an effective way.

Traditional cache replacement policies are designed for systems using DRAM memory. As DRAM doesn't have the problem of write endurance. These policies give no special considerations to reducing write traffic. The basic idea of cache replacement policies proposed for systems using PCM memory is preferentially keeping dirty blocks in cache under the premise of ensuring high hit ratio, because write operations to memory only occur when dirty blocks are evicted from LLC. A common defect of the existing works is that they don't distinguish the dirty block thoughout the whole lifetime of a cache block. The most obvious part is the demotion process (i.e., changing a long-term idle cache block to a relative low protection priority). Thus although they do give some special considerations to dirty cache blocks, the effects are limited. There are also some specialized page replacement policies proposed for systems using NAND flash [14] disk. Their design principles are instructive to us. As shown in Table I, NAND Flash also has the problem of write endurance. Due to its long latency, NAND Flash is commonly used as disk.

Table 1. The comparison of different memory technologies.

| Technology | DRAM | PCM | NAND FLASH |
|---|---|---|---|
| **Density** | 4X | <16X | 16X |
| **Read Latency** | 60ns | 200-300ns | 25us |
| **Write Speed** | ≈1Gbps | ≈100MB/s | 2.4MB/s |
| **Dynamic Power** | Medium | Medium read<br>High write | Very high read<br>Even high write |
| **Leak Power** | Medium | Low | Low |
| **Non-volatile** | No | Yes | Yes |
| **Scalability** | Yes | Yes | Yes |
| **Endurance** | N/A | $10^8 \sim 10^9$ | $10^6$ |
| **Retention** | N/A | 10 years | 10 years |

In this article, we propose a systematically multilevel structure considering both aspects of hit ratio and write traffic. This structure makes clear distinctions between cache blocks of different dirty degrees. To be brief, it divides the cache blocks into several priority levels taking both hit ratio and write traffic into account. Each level corresponds to a single type of cache blocks. The types of same dirty degree can be organized into one group. A cache block can only be changed to the priority levels within its group unless its dirty degree has changed. Then we design a novel replacement policy called MAC (Multilayer Ark for Cache) revolving around this hierarchy to protect the dirty blocks throughout their whole lifetime in cache. We evaluate the performance of MAC using GEM5 with benchmarks from PARSEC and SPLASH-2. We also select representatives of traditional cache replacement policies and most recently published works for PCM as contrasts. For convenience, we select LRU as baseline and normalize the original results. The normalized results show that RRIP, DRRIPW, LDF, RWA and MAC can averagely reduce the amount of writes to main memory by 4.68%, 10.39%, 6.2%, 17.62% and 25.12% respectively. While their differences in hit ratio and execution time are negligible.

Our work mainly makes the following two contributions:

*1) We analyse the essential factors of a cache replacement policy designed for systems using PCM memory, and construct a systematically multilevel structure to strictly make distinctions between cache blocks of different dirty degrees. This hierarchical structure considers both the aspects of write traffic and hit ratio. It separately divides cache blocks of different dirty degrees into several priority levels. A group corresponding to a larger dirty degree generally consists of higher priority levels, and a cache block can never be changed to the levels of other groups unless its dirty degree has changed.*

*2) Based on the structure above, we design a novel cache replacement policy called MAC. The whole policy consists of three parts: the insertion policy, the promotion policy and the victim selection policy. The insertion and promotion policies all distinguishes dirty and clean cache blocks (we don't make finer distinctions among dirty ones). The victim selection policy dutifully reflects the order of priority and make different demotion disposals for dirty and clean cache blocks. MAC gives the dirty cache block extra protection throughout its whole lifetime. It can effectively reduce writes to PCM memory with low hardware overhead, and hardly degrading the program performance.*

We arrange the rest content of the article as follows: Section 2 introduces the related work includes kinds of technologies to solve the problem of high cost of write to PCM. We mainly give brief descriptions of the representatively existing cache replacement policies we select. Section 3 and Section 4 illustrate our systematically multilevel structure and the novel cache replacement policy respectively. Then Section 5 is the experimental methodology, including the results and analysis. We make a conclusion in Section 6 and section 7 is the acknowledgement.

## 2. RELATED WORK

There are many works proposed to handle the problem of write operations to PCM. We summarize them into four main categories, the cache replacement policy is one of them. The rest three categories are utilizing extra storages [4, 6, 7, 20, 21], omitting the unnecessary writes [6, 9, 10, 17, 24] and making the load uniformly distributed [6, 24]. The category of cache replacement policy can be almost perfectly incorporated with the technologies of other categories. Below we give a relatively detailed introduction to it.

Among traditional cache replacement policies, LRU [11] is the best representative since it is most commonly used. LRU assumes that a recently used cache block is more likely to be re-referenced. It uses a chain structure to record the order of last references of all the cache blocks. The first-of-chain called MRU is for the most recently used block, and the end-of-chain is called LRU. LRU works well only when the data access pattern has high temporal locality. For example, if there are many single-use data, LRU barely brings a hit. RRIP [12] can effectively handle this situation by giving cache blocks high protection priority only when they are re-referenced. As for systems using PCM memory, recently works include DRRIPW [22], LDF [8], RWA [15] and I-RWA [15]. DRRIPW refers to a series of replacement policies, which are all based on DRRIP [12]. The main modification is that instead of using cache hit ratio as the criterion, DRRIPW uses Set Dueling [16] to select the replacement policy which creates less writes to main memory. LDF makes use of the fine-grained dirtiness management. It divides a cache block (e.g., 4KB) into several sub-blocks (e.g., 256B), and stores the number of dirty sub-blocks for each block. When it is time to select a victim, it victimizes a block with the least number of dirty sub-blocks among the candidates. LDF is more like an upgrade patch to be integrated into other policies. In order to utilize it, we need other technologies so that only the dirty part of a cache block will be written to PCM. There is a critical point that ignored by both DRRIPW and LDF: the cause of dirty cache blocks, which means the write back operation. RWA notices this factor and distinguishes write operations (write back) from read operations (load and store). RWA is based on SRRIP [12]. It gives dirty cache block the safest position in the beginning by modifying the progress of insertion.

There will be troubles if the re-reference intervals of dirty blocks are too long. To solve this problem, I-RWA narrows the difference between read and write operations in the progress of insertion, and makes up in the process of promotion under hits. CCF-LRU (Cold-Clean-First LRU) [13] is an efficient page replacement policy designed for systems using NAND flash disk. Its main idea is using labels to divide the pages into four protection priorities.

## 3. SYSTEMATICALLY MULTILEVEL STRUCTURE

In the existing cache replacement policies for systems using PCM memory, the special treatments of dirty cache blocks don't cover each process of cache. In order to make distinctions between the dirty cache blocks and clean ones throughout their whole lifetime, we construct a systematically multilevel structure as a basis of designing novel cache replacement policy.

This structure considers both aspects of hit ratio and write traffic. In the purpose of reducing write traffic to PCM memory, we need to preferentially protect the dirty cache blocks, while to get a high hit ratio, we also need to keep the blocks which are likely to be re-referenced. We adopt classification to quantify the combination property of each cache block.

As for hit ratio, we divide the blocks into $N_1$ levels. An extreme case is making each block a single level, and in this case $N_1$ reaches its maximum (e.g., The LRU cache replacement policy), so the value range of $N_1$ is as shown in (1). FL (Fresh Level) indicates which one of the $N_1$ levels the block belongs to, and its value range is as shown in (2).

$$1 \leq N_1 \leq \text{The Number of Cache Way} \qquad (1)$$

$$1 \leq \text{FL} \leq N_1 \qquad (2)$$

For write traffic, the blocks are divided into $N_2$ levels according to how many dirty bytes they have. Theoretically the finest situation is tracking each byte and record it is dirty or not, and $N_2$ reaches its maximum in this case. Its value range is as shown in (3) (B is the total number of bytes in a cache block). DL (Dirty Level) indicates which one of the $N_2$ levels the block belongs to. DL can take values in the range as shown in (4).

$$1 \leq N_2 \leq B + 1 \qquad (3)$$

$$1 \leq \text{DL} \leq N_2 \qquad (4)$$

The function of $N_1$, $N_2$, FL and DL is helping designer gain clarity. Taking both aspects into account, we have $N_T$ levels as in (5), and we add a label FDL (Fresh Dirty Level) to each cache block to indicate which one of the total $N_T$ levels it belongs to. FDL is the combination of FL and DL and has a value range as shown in (6). The specific value of FDL is determined by FL and DL together. There will be a correspondence table of FL, DL and FDL in practice.

$$N_T = N_1 \times N_2 \qquad (5)$$

$$1 \leq \text{FDL} \leq N_T \qquad (6)$$

The value of FDL stands for the protection priority of the cache block. For convenience, we assume that a larger FDL means a lower protection priority (FL and DL adopt the same assumption). The FDL of a block is changeable. For example, if a block hasn't been referenced for a long time, its FDL should be changed to a larger value, or it has more dirty bytes, then its FDL should be decreased.

It should be noticed that $N_1$, $N_2$, $N_T$, FL, DL and FDL are all integers. Through controlling the value of $N_1$ and $N_2$, and arranging the correspondence among FL, DL and FDL, we can intentionally change the attentions to each aspect, then get cache replacement policies of various effects.

We divide the cache blocks into different groups by making the cache blocks whose FDLs correspond to a same DL belong to a same group. A group of smaller DL generally correspond to smaller FDLs. If a cache block only changes its FL, its FDL will still in the value range of original group. Through this division, we can make cache blocks of different dirty degrees separately change their protection priorities within their own ranges.

At last, our systematically multilevel structure can be summarized as follows:

This structure divides the whole cache blocks into $N_T$ levels considering both aspects of keeping high hit ratio and reducing write traffic to PCM memory. It adds a label called FDL to each cache block to indicate which level of the $N_T$ ones it belongs to. FDL stands for the protection priority of a cache block. Cache blocks of same dirty degree belong to one group. The FDL of a cache block can only vary within the value range of its group unless the dirty degree of the cache block has changed. Though this we can give the dirty cache blocks roundly protections to reduce the write traffic.

## 4. NOVEL CACHE REPLACEMENT POLICY

Based on the systematically multilevel structure, we design a novel cache replacement policy for systems using PCM memory. This policy called MAC (Multilayer Ark for Cache) can effectively reduce write traffic to PCM memory with low hardware overhead.

We want MAC can be used in general situation, which means no modification to the process of writing data to PCM memory, so we only distinguish dirty and clean cache blocks. In that case the value of $N_2$ is set as 2. We set the value of $N_1$ as 2 so that the structure can be simple, and add a global LRU structure of all cache blocks as a supplement. According to (3), we have 4 protection priority levels in total.

We consider that the aspect of hit ratio relatively needs more attention. So we make sure that the protection priorities of cache blocks with FLs equal to 1 are still higher than those with FLs equal to 2, as shown in Table 2. The value of DL is 2 means that the cache block is clean, while 1 means dirty. The cache blocks with FLs equal to 1 are referenced more recently than those with FLs equal to 2.

Table 2. The correspondence between FL, DL and FDL.

| FDL | 1 | 2 | 3 | 4 |
|---|---|---|---|---|
| FL  | 1 | 1 | 2 | 2 |
| DL  | 1 | 2 | 1 | 2 |

We decompose the concrete content of MAC into three parts. The insertion policy is the method handling new cache blocks. The promotion policy presents the corresponding changes when a cache block is re-referenced. The victim policy is the algorithm of selecting a cache block to evict when necessary. The detailed introductions are as follows (read means load/store operation, and write means write back operation):

1) The Insertion Policy: In order to stop single-use data occupying cache, the FLs of all new blocks are set as 2 (the LRU chain can take care of the temporal locality). If a read miss occurs the new block is clean, which means its DL is 2. So as shown in Table 3, it is labelled with FDL equal to 4. If the new block is caused by a write miss, its DL is 1 and its FDL is 3. Under both situations, the new cache block will be inserted at the MRU position of the global LRU chain.

2) The Promotion Policy: When a cache block is re-referenced, its FL will be changed to 1, so FDL will be 1 or 2. If its DL is 1 (FDL is 1 or 3) before the hit, or 2 (FDL is 2 or 4) and under a write hit, in both situations its new FDL is 1, otherwise is 2. To visually view the processes, we show them in Table 3. Besides the change of FDL, the cache block being re-referenced will be promoted to the MRU position of the global LRU chain.

Table 3. The process of changing FDL under different hits.

| **FDL** | 1 | 2 | 3 | 4 |
| --- | --- | --- | --- | --- |
| **Read hit** | 1 | 2 | 1 | 2 |
| **Write hit** | 1 | 1 | 1 | 1 |

3) The Victim Policy: Though the global LRU chain, we can get the sub LRU chain of each level. The victim is selected following the steps below:

  a. Check if there is at least one level 4 (i.e., FDL is 4) block. If any, choose the LRU block among them as the victim. Otherwise go to step b.

  b. If there is any level 3 (i.e., FDL is 3) block, victimize the LRU block of them, meanwhile put the LRU block of level 2 (i.e., FDL is 2) blocks to the MRU position of the global LRU chain and change its FDL to 4. Then make the LRU block of level 1 (i.e., FDL is 1) blocks the new MRU block of the global LRU chain and change its FDL to 3. Otherwise go to step c.

  c. Check the level 2 (i.e., FDL is 2), if not empty, make their LRU block the victim. And the LRU block of level 1 (i.e., FDL is 1) will be moved to the MRU position of the global LRU chain with its FDL changed to 3. Otherwise go to step d.

  d. It is apparent that all the blocks in cache belong to level 1 (i.e., FDL is 1), so directly evict the LRU block.

We use pseudo-code to explicate the select-victim algorithm in Figure 1.

```
Algorithm Select-Victim
Input: G_LRU = global LRU chain
       FDL of each block
Output: the cache block to be evicted

    LRU_3 = the LRU block of all level 4 (i.e., FDL = 4) blocks
    LRU_2 = the LRU block of all level 3 (i.e., FDL = 3) blocks
    LRU_1 = the LRU block of all level 2 (i.e., FDL = 2) blocks
    LRU_0 = the LRU block of all level 1 (i.e., FDL = 1) blocks

    while (true)
        if (LRU_3 is not NULL)
            victim = LRU_3
            return victim
        else if (LRU_2 is not NULL)
            victim = LRU_2
            change FDL of LRU_1 to 3
            move LRU_1 to G_LRU's MRU position
            change FDL of LRU_0 to 2
            move LRU_0 to G_LRU's MRU position
            return victim
        else if (LRU_1 is not NULL)
            victim = LRU_1
            change FDL of LRU_0 to 2
            move LRU_0 to G_LRU's MRU position
            return victim
        else
            victim = LRU_0
            return victim
        end if
    end while
End Select-Victim
```

Figure 1. The pseudo-code of select-victim algorithm

As to analyse the hardware overhead of MAC, we use the case of 16-way cache as example. In order to construct the global LRU chain, MAC uses 4 bits for each cache block to record its position in the global LRU chain. Then to store the value of FDL, extra 2 bits are necessary. For each value of FDL, MAC needs a 4-bit register to track the LRU block of the set. We assume that the cache line size is 64 B, then in one cache set, compared with LRU the proportion of incremental bits of MAC is as shown in (7). The result is acceptable in practice.

$$\frac{MAC\ cache\ set\ size - LRU\ cache\ set\ size}{LRU\ cache\ set\ size} \times 100\% = \frac{6}{64 \times 16 + 8} \times 100\% \approx 0.58\% \tag{7}$$

## 5. PERFORMANCE EVALUATION

### 5.1. Experiment Setup

To evaluate the performances of MAC and other representative cache replacement policies, we conduct a series of experiments on GEM5 [23]. GEM5 is a powerful simulation tool encompassing system-level architecture as well as processor microarchitecture. The simulated system has eight cores of timing CPU modal using the X86 instruction set. The L1 cache is private.

The L2 cache and PCM are shared by all eight cores. We simulate detailed memory controllers as well. The parameter values are chosen in-line with other researcher's works. Table 4 shows the specific parameters of configurations.

We evaluate the performance of LRU, RRIP, DRRIPW, LDF and RWA for comparison. In order to display obvious differences between these cache replacement policies, we prefer to use test programs which have large amounts of data accesses so that there will be plenty of cache blocks evicted from the L2. The test programs are selected from SPLASH-2 [18] and PASEC [19]. The parameters of these programs are shown in Table 5.

Table 4. The parameters of baseline configuration.

| Component | Parameter |
|---|---|
| **Processing core** | 8-core/X86/timing |
| **L1 cache** | I-cache and D-cache each 32KB, 2-way, 64B linesize, writeback policy, LRU, 2 cycles |
| **L2 cache** | 512KB, 8 banks, 16-way, 64B linesize, writeback policy, 15 cycles |
| **PCM memory** | 8GB, 1024 cycles read latency, 4096 cycles write latency, 4KB page size |
| **Network configuration** | Crossbar, 1-cycle hop latency |

Table 5. The parameters of test programs.

| Workload | Program size |
|---|---|
| **FFT** | 256k points |
| **OCEAN** | 514×514 ocean |
| **RADIX** | 1048576 keys, radix=1024 |
| **BLACKSCHOLES** | simsmall |
| **CANNEAL** | simsmall |
| **DEDUP** | simsmall |
| **FLUIDANIMATE** | simsmall |
| **FREQMINE** | simsmall |
| **STREAMCLUSET** | simsmall |
| **SWAPTIONS** | simsmall |

| X264 | simsmall |
|---|---|

## 5.2. Result and Analysis

We use three performance criteria, which are writes (i.e., the amount of writeback operations to the PCM memory), hit ratio and execution time. Since the original results are not convenient for comparison, we select the system using LRU as baseline and carry out normalization. Actually in our experiments RRIP means SRRIP, and LDF means SRRIP integrated with LDF. The RRPVs of RRIP, DRRIPW, LDF and RWA are all 2 bits.

Figure 2 shows the comparison of writes. We can see that RRIP hardly provide obvious reduction (except in *fluidanimate*), sometimes it even brings more writes than LRU (e.g., in *streamcluster* and x264). The performance of LDF is close to RRIP, because the amount of writes only depends on how many dirty cache blocks are evicted, the effect of LDF is limited to distinguishing dirty cache blocks and clean ones for RRIP. In most cases, DRRIPW and RWA produce less writes, and by contrast RWA performs better than DRRIPW (except in *radix*). Compared with the others, MAC presents a better performance in almost each program. This proves that MAC works effectively in various situations. In some programs (i.e., *swaptions*. *fluidanimate* and *blacksholes*) it brings a significant more than 50% reduction.

Table 6 lists the average results. RRIP, DRRIPW, LDF, RWA and MAC averagely reduce the writes to PCM memory by 4.68%, 10.39%, 6.2%, 17.62% and 25.12% respectively compared with LRU. The effects on reducing writes of DRRIPW, RWA and MAC are considerable. Both RWA and MAC consider the difference between write and read operations, thus they all perform better than DRRIPW. By distinguishing write operation from read operation, RWA and MAC can recognize the dirty cache blocks in the process of inserting new blocks into cache, then take corresponding protective measure. For example, when a write miss occurs, RWA and MAC will insert the new cache block in a relatively higher protection level since it will be dirty after the write, while DRRIPW just treats it in the same way of the clean cache block following a read miss. MAC occurs less writes than RWA because it protects dirty cache blocks in the process of demotion while RWA doesn't. For example, we use numbers from 1 to 4 to denote the protection priority level, and 1 means safest. If a dirty cache block hasn't been re-referenced for a long time, it can be downgraded to level 4 according to RWA, which means most likely to be evicted. While in MAC, dirty cache blocks can only be level 1 or 3. So no matter how long a dirty cache block has been idle, it is totally safe as long as there is one clean cache block in level 4.

We divide the whole lifetime of a cache block in to four process:

1) Insertion: the process of inserting the cache block into cache.

2) Promotion: the process of giving the cache block higher protection priority when it is re-referenced.

3) Demotion: the process of giving the cache block lower protection priority if it is long-term idle.

4) Selection: the process of deciding whether if evict the cache block.

Table 7 shows that whether DRRIPW, LDF, RWA and MAC distinguish dirty cache block or not in each process. We can see that only MAC covers all the four processes.

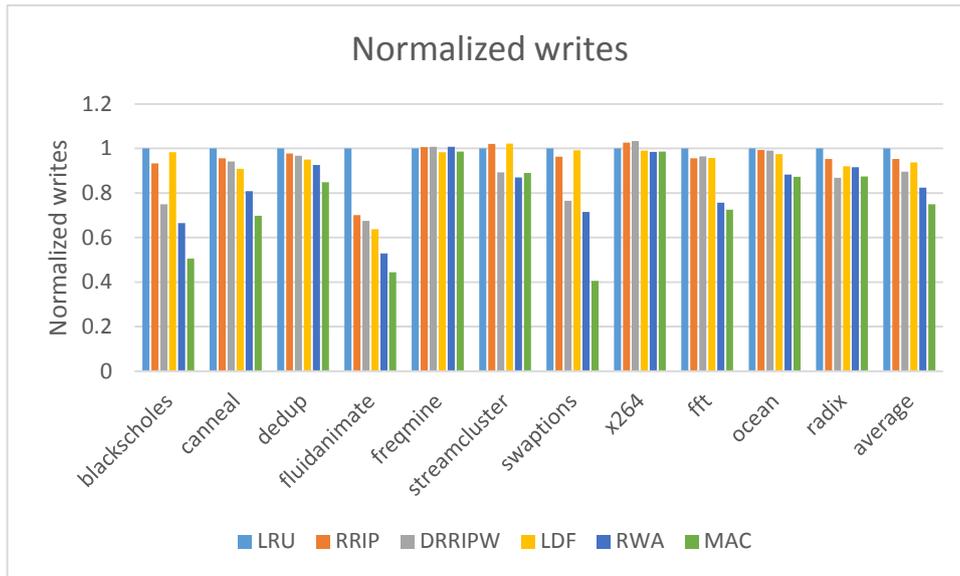

Figure 2. The results of writes (normalized to LRU)

Table 6. The average results of writes (normalized to LRU).

| Replacement policy | Average results of writes |
|---|---|
| **LRU** | 100% |
| **RRIP** | 95.32% |
| **DRRIPW10** | 86.61% |
| **LDF** | 93.80% |
| **RWA** | 82.38% |
| **MAC** | 74.88% |

Table 7. Whether distinguish dirty block in each process. (✓ means yes and ✗ means no)

| Replacement policy | Insertion | Promotion | Demotion | Selection |
|---|---|---|---|---|
| **DRRIPW10** | ✗ | ✓ | ✗ | ✓ |
| **LDF** | ✗ | ✗ | ✗ | ✓ |
| **RWA** | ✓ | ✓ | ✗ | ✓ |
| **MAC** | ✓ | ✓ | ✓ | ✓ |

As shown in Figure 3, in most programs the differences of hit ratio between these policies are negligible. The average results are in Table 8. Compared with LRU, RRIP averagely increases the hit ratio by 1.41% due to the single-use data. LDF also increases by 1.15% since its hit ratio is determined by the cache replacement policy it integrated with. DRRIPW, RWA and MAC respectively decrease the hit ratio by 1%, 0.72% and 0.41% in average. They spare part attentions to reducing the writes, sometimes a dirty cache block with slightly low probability of being re-referenced is more worth preserving than a clean one with high probability of being re-referenced. Considering their contributions of reducing writes, the influences of hit ratio are acceptable. Among them MAC has the lowest degradation of hit ratio.

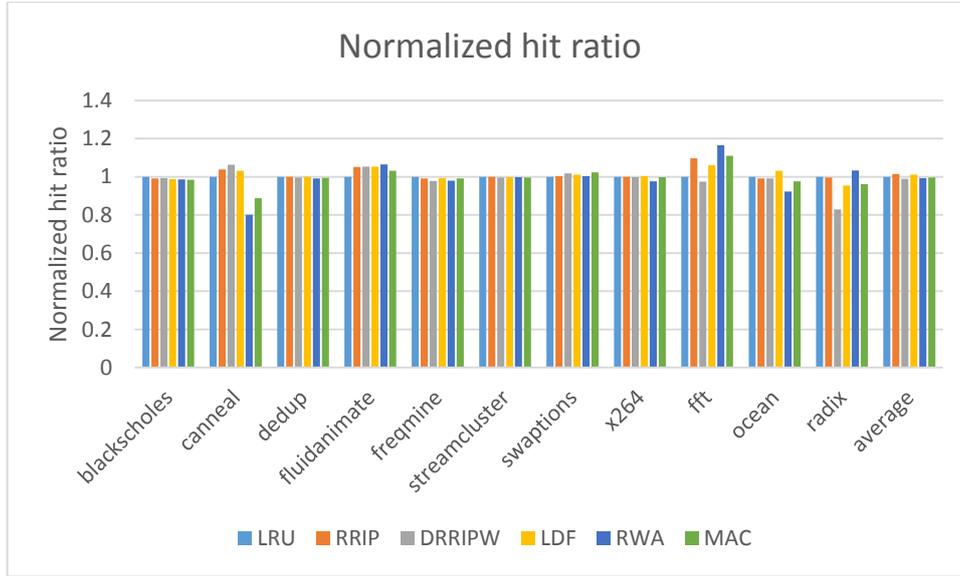

Figure 3. The results of hit ratio (normalized to LRU)

Table 8. The average results of hit ratio (normalized to LRU).

| Replacement policy | Average results of hit ratio |
| --- | --- |
| **LRU** | 100% |
| **RRIP** | 101.41% |
| **DRRIPW10** | 99.00% |
| **LDF** | 101.15% |
| **RWA** | 99.28% |
| **MAC** | 99.59% |

Figure 4 shows the results of execution time. Except in *freqmine* and *streamcluster*, the performances of these policies are almost the same. As shown in Table 9, RRIP, DRRIPW, LDF and MAC shorten the execution time by 0.32%, 0.42%, 0.23% and 0.44% respectively in comparison to LRU. As to PCM, the latency of a write operation may be longer than the total latency of several read operations. Although DRRIPW and MAC have relatively low hit ratios, their execution times are all shorter than LRU due to less write operations.

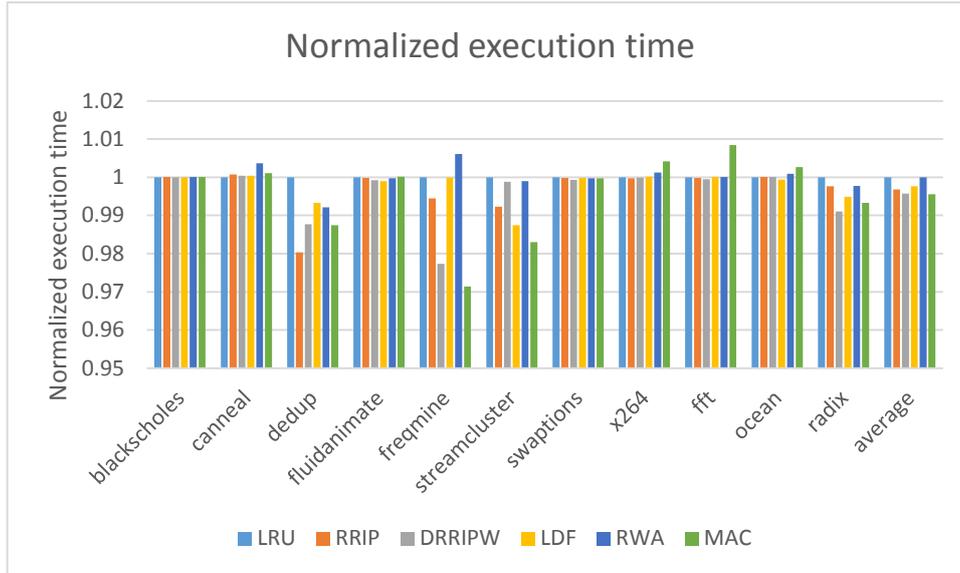

Figure 4. The results of execution time (normalized to LRU)

Table 9. The average results of execution time (normalized to LRU).

| Replacement policy | Average results of execution time |
| --- | --- |
| **LRU** | 100% |
| **RRIP** | 99.68% |
| **DRRIPW10** | 99.58% |
| **LDF** | 99.77% |
| **RWA** | 100.00% |
| **MAC** | 99.56% |

## 6. CONCLUSIONS

In this paper we proposed a systematically multilevel structure for designing cache replacement policy to reduce write traffic to PCM memory, and based on the structure we present an effective cache replacement policy called MAC. MAC can greatly reduce the amount of writes to PCM with low hardware overhead, and hardly brings degradation of program performance.

Through the simulation experiments we proved that MAC is more efficient in the aspect of reducing writes compared with LRU, RRIP, DRRIPW, LDF and RWA. In some programs (e.g., *swaptions* and *fluidanimate*) MAC achieves dramatically more than 50% reductions compared with LRU. The average results show that MAC produces the least writes and shortest execution time among these policies.

## ACKNOWLEDGEMENTS

This work is supported by SRFDP (Grant No 20120002110032) and NSF of China (Grant No 61303002, 61373025).

**REFERENCES**

[1] Lefurgy C, Rajamani K, Rawson F, et al. Energy management for commercial servers[J]. Computer, 2003, 36(12): 39-48.

[2] Wilson L. International Technology Roadmap for Semiconductors (ITRS)[J]. Semiconductor Industry Association, 2013.

[3] Li J, Lam C. Phase change memory[J]. Science China Information Sciences, 2011, 54(5): 1061-1072.

[4] Lee B C, Zhou P, Yang J, et al. Phase-change technology and the future of main memory[J]. IEEE micro, 2010, 30(1): 143.

[5] J. Coburn, A. M. Caulfield, A. Akel, L. M. Grupp, R. K. Gupta, R. Jhala, and S. Swanson. NV-Heaps: making persistent objects fast and safe with next-generation, non-volatile memories. In International Conference on Architectural Support for Programming Languages and Operating Systems, 2011.

[6] Qureshi M K, Srinivasan V, Rivers J A. Scalable high performance main memory system using phase-change memory technology[J]. ACM SIGARCH Computer Architecture News, 2009, 37(3): 24-33.

[7] Wu X, Li J, Zhang L, et al. Hybrid cache architecture with disparate memory technologies[C]//ACM SIGARCH computer architecture news. ACM, 2009, 37(3): 34-45.

[8] Yoo S, Lee E, Bahn H. The least-dirty-first cache replacement policy for phase-change memory[C]//Proceedings of the 29th Annual ACM Symposium on Applied Computing. ACM, 2014: 1449-1454.

[9] Zhou P, Zhao B, Yang J, et al. A durable and energy efficient main memory using phase change memory technology[C]//ACM SIGARCH Computer Architecture News. ACM, 2009, 37(3): 14-23.

[10] Cho S, Lee H. Flip-n-write: a simple deterministic technique to improve pram write performance, energy and endurance[C]//Microarchitecture, 2009. MICRO-42. 42nd Annual IEEE/ACM International Symposium on. IEEE, 2009: 347-357.

[11] Lilly B P, Williams III G R, Sadoughi-yarandi M, et al. Least Recently Used Mechanism for Cache Line Eviction from a Cache Memory: U.S. Patent 20,150,026,404[P]. 2015-1-22.

[12] Jaleel A, Theobald K B, Steely Jr S C, et al. High performance cache replacement using re-reference interval prediction (RRIP)[C]//ACM SIGARCH Computer Architecture News. ACM, 2010, 38(3): 60-71.

[13] Li Z, Jin P, Su X, et al. CCF-LRU: a new buffer replacement algorithm for flash memory[J]. Consumer Electronics, IEEE Transactions on, 2009, 55(3): 1351-1359.

[14] Maejima H, Isobe K. NAND flash memory: U.S. Patent 8,630,116[P]. 2014-1-14.

[15] Zhang X, Hu Q, Wang D, et al. A read-write aware replacement policy for phase change memory[M]//Advanced Parallel Processing Technologies. Springer Berlin Heidelberg, 2011: 31-45.

[16] Qureshi M K, Jaleel A, Patt Y N, et al. Adaptive insertion policies for high performance caching[C]//ACM SIGARCH Computer Architecture News. ACM, 2007, 35(2): 381-391.

[17] Fang Y, Li H, Li X. SoftPCM: Enhancing energy efficiency and lifetime of phase change memory in video applications via approximate write[C]//Test Symposium (ATS), 2012 IEEE 21st Asian. IEEE, 2012: 131-136.

[18] Woo S C, Ohara M, Torrie E, et al. The SPLASH-2 programs: Characterization and methodological considerations[C]//ACM SIGARCH Computer Architecture News. ACM, 1995, 23(2): 24-36.

[19] Bienia C, Kumar S, Singh J P, et al. The PARSEC benchmark suite: Characterization and architectural implications[C]//Proceedings of the 17th international conference on Parallel architectures and compilation techniques. ACM, 2008: 72-81.

[20] Lee H G, Baek S, Nicopoulos C, et al. An energy-and performance-aware DRAM cache architecture for hybrid DRAM/PCM main memory systems[C]//Computer Design (ICCD), 2011 IEEE 29th International Conference on. IEEE, 2011: 381-387.

[21] Ramos L E, Gorbatov E, Bianchini R. Page placement in hybrid memory systems[C]//Proceedings of the international conference on Supercomputing. ACM, 2011: 85-95.

[22] Rodrǵuez-Rodrǵuez R, Castro F, Chaver D, et al. Reducing writes in phase-change memory environments by using efficient cache replacement policies[C]//Proceedings of the Conference on Design, Automation and Test in Europe. EDA Consortium, 2013: 93-96

[23] The gem5 Simulator, http://www.m5sim.org/.

[24] Hu J, Xue C J, Zhuge Q, et al. Write activity reduction on non-volatile main memories for embedded chip multiprocessors[J]. ACM Transactions on Embedded Computing Systems (TECS), 2013, 12(3): 77.

**Authors**

Shenchen Ruan is a master candidate of Tsinghua University, and majors in computer science. He received a bachelor's degree from Tsinghua University. He is a member of the research centre of CPU, and his major research direction is about computer architecture.

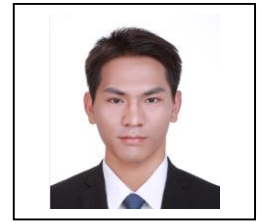

Haixia Wang is an associate professor at Tsinghua University. She holds a doctor's degree and in charge of a group in the research centre of CPU. The group mainly focuses on the reach of optimizing computer system architecture to meet different application requirements (e.g., handling large data sets).

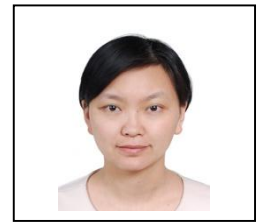

Dongsheng Wang is a professor at Tsinghua University, and the director of the research centre of CPU. He has a doctor's degree. He won the prize of "Ten outstanding postdoctoral" during his time in the postdoctoral workstation at Tsinghua University, when he worked on the research of parallel processing and distributed computer system, high availability and fault tolerant technology.

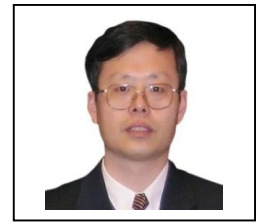